%Paper: hep-ph/9406244
%From: mikeb@a3.ph.man.ac.uk
%Date: Tue, 7 Jun 1994 15:53:16 +0100

\magnification\magstep 1
\vsize=21 true cm
\hsize=16.7 true cm
\baselineskip=0.65 true cm
\parindent=1.1 true cm

\raggedbottom

\font\bfmag=cmbx10 scaled\magstep3
\font\bfmagd=cmbx10 scaled\magstep2
\font\bfmagc=cmbx10 scaled\magstep1

\font\ennearm=cmr9

{\rightline {MC/TH 94/10}}

\bigskip

\bigskip

\bigskip

\centerline {\bfmag Constraints on Four-Point Couplings}
\centerline {\bfmag in Low-Energy Meson Interactions}
\bigskip
\bigskip
\centerline{\bfmagc Dimitri Kalafatis and Michael C. Birse}
\bigskip
\centerline{\sl Theoretical Physics Group}
\centerline{\sl Department of Physics and Astronomy}
\centerline{\sl The University of Manchester}
\centerline{\sl Manchester M13 9PL}
\centerline{\sl UNITED KINGDOM}

\bigskip

\bigskip

\centerline{June 1994}

\bigskip

\bigskip

\centerline{\bfmagc Abstract}

{\ennearm We investigate from first principles the introduction of isospin-1
vector and axial-vector fields into the nonlinear sigma model. Chiral symmetry
is nonlinearly realised and spin-1 fields are assumed to transform
homogeneously under chiral rotations. By requiring the Hamiltonian of the
theory to be bounded from below we find inequalities relating three- and
four-point meson couplings. This leads to a low-energy phenomenological
Lagrangian for the nonanomalous sector of $\pi\rho a_1$ strong interactions}.

\bigskip

\bigskip

\centerline{\bfmagc{1. Introduction}}

\medskip

At low energies strong interactions can be described by an effective Lagrangian
in terms of mesons [1]. This should comply with the approximate symmetries of
low energy strong interactions such as chiral symmetry. Chiral symmetry
provides information about the general structure of the couplings between
mesons whereas the coupling constants entering the effective Lagrangian are
related to more detailed features of the underlying QCD dynamics. Unfortunately
it is still impossible to extract from QCD the values of the effective low
energy parameters. Some of these can be related to phenomenologically known
meson observables, like masses and decay widths, but most of the higher order
parameters remain unknown.

The starting point for such an effective Lagrangian is the nonlinear sigma
model of pseudoscalar pions. This realises spontaneous breaking of chiral
symmetry, a central feature of low energy QCD and introduces a single
parameter, the pion decay constant. The experimental discovery of meson
resonances as well as some theoretical notions such as the large $N_c$
expansion of QCD [2], strongly support the idea of introducing mesons other
than the pion into this model. There is a considerable amount of work in the
literature treating the role played by massive spin-$1$ mesons (the $\rho$- and
the $a_1$-mesons) in low-energy Lagrangians. In most of these works isovector
resonances are introduced as massive Yang-Mills particles [3] or as gauge
bosons of local chiral symmetry [4]. In these approaches some of the new
coupling constants can be determined by fitting to processes like
$\rho\to\pi\pi$ or $a_1\to\rho\pi$. The remaining four-point and three-point
coupling constants are then completely determined by the gauge symmetry
assumption.

Although these approaches are consistent with the phenomenologically successful
notion of vector meson dominance, it should be noted that there is neither
experimental evidence nor theoretical prejudice from QCD to support the
existence of a gauge symmetry in low energy hadronic interactions. Furthermore
as the authors of Ref. [5] have shown vector meson dominance is not a feature
unique to the models of Refs. [3-4]. It can also be obtained in models where
chiral symmetry is realised in a different manner.

Here we follow a different approach, as suggested by [6], of writing down a
general Lagrangian consistent with basic principles of field theory and chiral
symmetry. Vector meson dominance can be implemented later, if so desired, by
specific choices of parameters. From this point of view it is reasonable to
assume a homogeneous transformation law for isovector spin-1 fields, instead of
that used in ref.~[4]. As was shown recently for the case of the $\pi\rho$
system [6], without making any additional symmetry assumption, constraints
relating three- and four-point coupling strengths do exist. These derive from
demanding the Hamiltonian to be bounded from below. The results of [6]
are encouraging enough to suggest a systematic investigation of four-point
couplings in more realistic theories that include the axial-vector meson.

The purpose of this work is therefore to extend the analysis of ref.~[6] to the
description of interacting pions, $\rho$- and $a_1$-mesons assuming that the
spin-1 isovector fields transform homogeneously under nonlinear chiral
symmetry. The $\pi\rho a_1$ system turns out to be more complicated than the
$\pi\rho$ case but it is more interesting since there are both  vector and
axial-vector mesons with masses of around $1$ GeV. In section 2 we define the
transformation properties of the fields. Section 3 is devoted to the
investigation of the energies of nonperturbative field configurations in the
framework of the minimal three-point coupling theory.  We show that these
energies are unbounded from below. In section 4 we show that the inclusion of
four-point effective couplings counterbalances the dangerous contributions of
the three-point terms to these energies. We derive inequalities between three-
and four- meson couplings for the theory to make sense. In section 5 we discuss
how unitarity arguments based on vector dominance could lead to saturation of
these inequalities and present a low-energy $\pi\rho a_1$ effective Lagrangian
consistent with chiral symmetry and general field theoretical principles.

\bigskip

\bigskip

\centerline{\bfmagc{2. Transformations under chiral rotations}}

\medskip

Our starting point is the Lagrangian of the nonlinear sigma model defined in
terms of the $SU(2)$ field $U$ as:
$$\eqalign{
{\cal{L}}_{NL\sigma}={{f^2}\over 4}<\partial^\mu U \partial_\mu
U^\dagger >,\cr}\eqno(1)$$
$f$ being the pion decay constant and the symbols \lq\lq $< \ >$" denoting a
trace in SU(2) space. Since we are interested here in the structure of the
theory for large amplitude field configurations we define $U$ as
$U=\exp{(i\vec\tau.\vec F(x))}$ with the pion field given by $\vec F=F \hat F$.
Other parametrisations are perhaps more suitable for perturbative evaluations
of Green' s functions, but are not as convenient for investigations of the
large field region.

The Lagrangian of eq.~(1) is manifestly invariant under the linear
$SU(2)_L\otimes SU(2)_R$ global transformation $U\to g_L U g_R^\dagger$ with
$g_L,g_R \in SU(2)$. It is also invariant under the following nonlinear
rotation [7] of the square root $u$ of $U$:
$$\eqalign{
u(\vec F) \ \to \ g_L u(\vec F) h^\dagger(\vec F) \ = \ h(\vec F)
u(\vec F) g_R^\dagger,\cr}
\eqno(2)
$$
$h(\vec F)$ being an $SU(2)$-matrix that depends nonlinearly on the pion
fields. This compensating
transformation $h(\vec F)$ ensures that $U$ transforms linearly.

With the pion unitary matrix transforming as in eq.~(2)
one defines the following field gradients:
$$\eqalign{
u_\mu=&i (u^\dagger\partial_\mu u-u\partial_\mu u^\dagger)\cr
\Gamma_\mu=&{1\over 2}(u^\dagger\partial_\mu u+u\partial_\mu u^\dagger).\cr}
\eqno(3)
$$
The axial-vector and vector characters respectively
of $u_{\mu}$ and $\Gamma_{\mu}$ are manifest from their expressions
in terms of pseudoscalar pion fields $\vec F$:
$$\eqalign{
u_\mu=&-\tau_k \ \big[\hat F_k\hat F_m+{\displaystyle{\sin F
\over F}} (\delta_{km}-\hat F_k\hat F_m)\big] \ \partial_\mu
\vec F_m\cr
\Gamma_\mu=& \ i \vec\tau \ . \ (\vec F\times
\partial_\mu\vec F)  \ {\displaystyle{{\sin^2(F/2)}\over
{F^2}}}.\cr}\eqno(4)
$$
{}From eq.~(2) the transformations of these
gradients under chiral symmetry are given by:
$$\eqalign{
u_\mu \ \to& \ h(\vec F) u_\mu h^\dagger (\vec F) \cr
\Gamma_\mu \ \to& \ h(\vec F) \Gamma_\mu h^\dagger (\vec F) +
h(\vec F) \partial_\mu h^\dagger (\vec F).\cr}\eqno(5)
$$
The quantity $u_\mu$ is seen to transform
homogeneously whereas the transformation of
$\Gamma_\mu$ contains an inhomogeneous part as a result of
the field dependence of $h(\vec F)$.

In extending this to spin-1 isovector particles, in particular the $\rho$ and
the $a_1$, the immediate question is: how should these fields transform in this
framework? Using the matrix $h$ one finds that if these fields are to be
described by Lorentz vectors there are only two possibilities forming a group:
homogeneous or inhomogeneous.

In the case of inhomogeneous transformation laws [4] the associated lowest
order invariant Lagrangian preserves not only chiral symmetry but also a
certain sort of a gauge symmetry. Furthermore in the inhomogeneous approach it
is impossible to define similar transformations for both the $\rho$ and the
$a_1$ fields, simply because the associated particles have opposite parity.

In contrast the homogeneous transformation is the simplest one consistent with
chiral symmetry and has the nice feature that it can be applied to both the
$\rho$ and the $a_1$ fields. We adopt this first-principles point of view and
assume that the $\rho$- and the $a_1$-mesons transform homogeneously under the
nonlinear chiral group:
$$\eqalign{
V_\mu \ \to &\ h(\vec F) V_\mu h^\dagger(\vec F)\cr
A_\mu \ \to &\ h(\vec F) A_\mu h^\dagger(\vec F)\cr}\eqno(6)
$$
where $V_\mu=\vec\tau.\vec V_\mu$ and $A_\mu=\vec\tau.\vec A_\mu$. It is clear
that $\Gamma_\mu$ is the necessary ingredient for the definition of covariant
derivatives of spin-1 fields transforming as in eq.~(6)
$$\eqalign{
\nabla_{\mu} =\partial_{\mu} +
[\Gamma_\mu, \ ].\cr}\eqno(7)
$$
It is easy to check now that $\nabla_\mu V_\nu$ and $\nabla_\mu
A_\nu$ also transform homogeneously: $\nabla_\mu V_\nu\to
h\nabla_\mu V_\nu\ h^\dagger$ and similarly $\nabla_\mu A_\nu\to
h \nabla_\mu A_\nu\ h^\dagger$.

\bigskip

\bigskip

\centerline{\bfmagc{3. Three-point couplings}}

\medskip

With the transformation rules defined previously the invariant Lagrangian at
quadr\-atic order in the fields is given by
$$\eqalign{
{\cal{L}}_{\pi\rho a_1}^{(2)}=&{{f^2}\over 4}<u_\mu u^\mu>-
{1\over 4}<V_{\mu\nu}V^{\mu\nu}>
-{1\over 4}<A_{\mu\nu}A^{\mu\nu}>\cr
&+{{M_{\rho}^2}\over 2}<V_\mu V^\mu>+
{{M_a^2}\over 2}<A_\mu A^\mu>,\cr}\eqno(8)
$$
where $V_{\mu\nu}=\nabla_\mu V_\nu-\nabla_\nu V_\mu$ and $A_{\mu\nu}=\nabla_\mu
A_\nu-\nabla_\nu A_\mu$ are the covariant field strengths of the spin-1
resonances. We introduce chirally invariant mass terms for the $\rho$- and the
$a_1$-mesons and we assume that the coupling $c< A_\mu u^\mu >$ is not present.
This latter coupling is a result of $a_1-\pi$ mixing which, at lowest order, is
certainly not allowed if at some level one is to identify the fields with the
physical states. With the choice $c=0$ no diagonalisation of $\pi\rho a_1$
interactions is needed - obviously not a disadvantage of our framework.

At the three-point level there is a number of possible chirally invariant terms
consistent with charge conjugation and parity invariance. Leaving three-point
interactions among the vector mesons aside for a future analysis, we consider
here chirally invariant three-point couplings with at least one pion field
gradient:
$$\eqalign{
{\cal{L}}_{\pi\rho a_1}^{(3)}=-{i\over{2\sqrt{2}}}\bigg\{ \ &g_1 <V_{\mu\nu}
[u^\mu,u^\nu]>  \ +  \ g_2 <A_{\mu\nu}\big([V^\mu,u^\nu]-[V^\nu,u^\mu]\big)>\cr
+ &g_3 <V_{\mu\nu} \big([A^\mu,u^\nu]-[A^\nu,u^\mu]\big)> \ \bigg\}.\cr}
\eqno(9)$$
The Lagrangian ${\cal{L}}_{\pi\rho a_1}^{(2)}+{\cal{L}}_{\pi\rho a_1}^{(3)}$
has six free parameters that can be determined by fitting to low energy meson
observables like masses, decay widths etc. In principle one would like to
determine these parameters from QCD but, while some recent investigations in
the ENJL model [8] suggest that the problem is not hopeless, a sensible method
to perform such an extraction from QCD has not yet been discovered.

The issue we address here is rather different: assuming that $g_1, g_2, g_3$
are somehow given by the underlying QCD dynamics, are there any relations
between these parameters and higher order ones? The results of ref.~[6] suggest
that this question should be addressed in a nonperturbative framework. In
particular does the theory defined by equations ~(8, 9) yield a Hamiltonian
that is bounded from below? To find an answer we study the effect of
three-point interactions in the classical sector of the theory. We construct
the Hamiltonian associated with the Lagrangian ${\cal{L}}_{\pi\rho
a_1}^{(2)}+{\cal{L}}_{\pi\rho a_1}^{(3)}$ in terms of the  canonical degrees of
freedom: the fields ${\vec F,\ \vec V_i,\ \vec A_i}$ and their conjugate
momenta, respectively ${\vec \phi,\ \vec \pi_i, \vec \chi_i}$. The Hamiltonian
functional can be written as a sum of two terms $H=H_T+H_V$, where the kinetic
energy is $H_T$ and the potential energy is $H_V$. The potential part contains
only space components and in the three-point case is given by  $$\eqalign{
H_V=&\int d^3 x \bigg\{ {{f^2}\over 2}(u_i)_k^2+M_a^2(A_i)_k^2 +{1\over
2}(A_{ij})_k\big[ \ A_{ij}+i\sqrt{2}g_2([V_i,u_j]-[V_j,u_i])\  \big]_k\cr
&+M_{\rho}^2(V_i)_k^2+{1\over 2}(V_{ij})_k\big[ \ V_{ij}+i\sqrt{2}g_1[u_i,u_j]
+i\sqrt{2}g_3([A_i,u_j]-[A_j,u_i]) \ \big]_k\bigg\}.\cr}\eqno(10) $$
The kinetic piece needs some work in order to eliminate the dependent variables
$\vec V_0,\ \vec A_0$. A detailed derivation of it is given in the Appendix;
here we simply state the result: $$\eqalign{ H_T=\int d^3 x \bigg\{ {1\over 2}
\vec\Phi   {\cal{A}}^{-1} \vec\Phi   +{{\vec\pi_i^2}\over 4} +
{{\vec\chi_i^2}\over 4} +{1\over 2} \vec\Gamma {\cal{P}}^{-1} \vec\Gamma
\bigg\},\cr}\eqno(11) $$ where $\vec\Phi,\ \vec\Gamma$ are linearly related to
the momenta  ${\vec \phi,\ \vec \pi_i,\ \vec \chi_i}$ and ${\cal{A}},\
{\cal{P}}$ are  isospin tensor functions of ${\vec F,\ \vec V_i,\ \vec A_i}$.
The rather lengthy expressions for these objects are also given in the
Appendix.

In order to exhibit the problematic structure of the theory defined by eq.~(9)
we investigate the energy of a classical coherent configuration of the meson
fields. The simplest such object one can imagine has an isospin content
specified by a constant unit vector $\hat F$:
$$\vec F_0(\vec x)= F(\vec x)\ \hat F, \eqno(12)$$
where $F(\vec x)$ is a regular function of space. Such a configuration is
topologically trivial and carries no baryon number. For the vector and the
axial vector fields it is convenient to assume that they are parallel to the
pion field:
$$\eqalign{
\vec V_i(\vec x)=& V_i(\vec x) \ \hat F \cr
\vec A_i(\vec x)=& A_i(\vec x) \ \hat F .\cr}\eqno(13)
$$
It is clear from the definitions in eqs.~(12-13) that all commutators between
space components including the connection $\Gamma_i$ vanish. As a result the
potential energy is simply given by:
$$\eqalign{
H_V=\int d^3 x \bigg\{
{{f^2}\over 2}(\partial_i F)^2+M_a^2A_i^2
+{1\over 2}(\partial_{i}A_{j}-\partial_j A_i)^2+M_{\rho}^2V_i^2+
{1\over 2}(\partial_{i}V_{j}-\partial_j V
_i)^2\bigg\},\cr}\eqno(14)
$$
the potential energy of a free theory as if $F$ was a massless
scalar field and $V_i$ and $A_i$ were the space components of two massive
spin-1 mesons. The important feature for our investigation is that this
potential energy is positive and does not depend on the couplings
of the theory. We therefore concentrate in the kinetic energy of
the theory as given by $H_T$.

To evaluate the kinetic energy for our meson state we need the explicit
structure of matrices ${\cal{A}}$ and ${\cal{P}}$ (see the Appendix) in terms
of the fields $F,\ V_i,\ A_i$. Because of the particular isospin structure we
consider here all these matrices can be simply decomposed in terms of two
symmetric isospin tensors: the unit tensor and $\hat F\otimes\hat F$. As a
consequence only momenta that point in a direction perpendicular to that of
the pion actually \lq\lq see" the couplings to the vector mesons. We assume the
following forms:
$$\eqalign{
\vec \phi=& \phi(\vec x) \ \hat \phi\cr
\vec \pi_i=& \pi_i(\vec x) \ \hat F \cr
\vec \chi_i=& \chi_i(\vec x) \ \hat F,\cr}\eqno(15)
$$
with $\hat\phi\cdot\hat F=0$. Using now the forms of eqs. (12, 13, 15)
in (11) we find the kinetic energy of our field configuration
$$\eqalign{
H_T=\int d^3 x \bigg\{{{\phi^2}\over{2f^2s^2 {\cal{I}}}}
+{1\over 4}\bigg[\pi_i^2+\chi_i^2+{{(\partial_i\pi_i)^2}\over{M_\rho^2}}
+{{(\partial_i\chi_i)^2}\over{M_a^2}}\bigg]\bigg\},\cr}\eqno(16)
$$
where $s$ is a shorthand notation for $\sin{F}/F$.

The crucial feature here is the structure of the dimensionless \lq\lq inertial"
parameter ${\cal{I}}$ which contains all the nontrivial effects due to the
inclusion of massive spin-1 fields:
$$\eqalign{
{\cal{I}}={1\over{f^4M_1M_2}}\bigg(&f^4M_1M_2-8(g_2^2 V_i^2M_2M_\rho^2
+(g_1\partial_i F-g_3 A_i)^2M_1M_a^2)\cr
&+16\big[g_2^4M_2(V_i^2(\partial_i F)^2-(V_i\partial_i F)^2)
+g_3^4M_1(A_i^2(\partial_i F)^2-(A_i\partial_i F)^2)\big]\bigg),\cr}\eqno(17)
$$
with $M_1=(1/f^2)\displaystyle{\big[2M_\rho^2-4g_2^2(\partial_i F)^2\big]}$ and
$M_2=(1/f^2)\displaystyle{\big[2M_a^2-4g_3^2(\partial_i F)^2\big]}$.
While in the case of the nonlinear sigma model with vanishing
couplings $g_1,\ g_2,\ g_3$ the \lq\lq inertial" parameter ${\cal{I}}$ is
simply equal to $1$, here it acquires negative contributions from the
vector and the axial-vector fields. For very small fields $F,\ V_i,\ A_i\approx
0$ appropriate to perturbation theory one has ${\cal{I}}\approx 1$ so the
problem does not appear in perturbative expansions of scattering amplitudes.

For nonperturbative configurations the situation changes dramatically since
then negative contributions proportional to quadratic powers of the couplings
can drive ${\cal{I}}$ to zero or negative values. The Hamiltonian density
acquires poles and the energy is not bounded from below. To give an idea of the
energetic scales where such troubles arise let us consider a localised meson
wave carrying momentum $k_i$ and of amplitude $F \approx 1$. As a further
simplification of our original ansatz we assume that all classical fields
vanish except $F$ and $\phi$. The gradient $\partial_i F$ is roughly
approximated by $k_i$ and ${\cal{I}}$ inside the meson wave looks like:
$$\eqalign{
{\cal{I}}\approx \displaystyle{{1-2\bigg(\displaystyle{{g_3^2}\over{M_a^2}}
+2\displaystyle{{g_1^2}\over{f^2}}\bigg)k^2}\over{
1-2\displaystyle{{g_3^2}\over{M_a^2}}k^2}}.\cr}\eqno(18)
$$
At small or very large momenta $k$ the inertial parameter is positive since the
denominator and numerator in eq.~(18) then have the same sign. But for $k^2$ in
the intermediate range
$$\eqalign{
\left(2{g_3^2\over M_a^2}+4{g_1^2\over f^2}\right)^{-1} < k^2 <
{M_a^2\over 2g_3^2}\cr}\eqno(19)
$$
${\cal{I}}$ becomes negative and as a consequence {\it the kinetic energy
density is negative}, making the theory ill defined in these regions.

Taking reasonable numerical values of the coupling constants [9], the region
of dangerous momenta is found to be $0.4$ GeV $< k < 2.0$ GeV, which
includes the range of masses of the $\rho$ and the $a_1$ resonances. However
this is precisely the range that one would like to describe by extending the
low-energy effective theories to include spin-1 mesons. Notice
also that if one switches off the couplings to the axial-vector meson,
$g_3=g_2=0$, the region where ${\cal{I}}$ becomes negative is modified to
$f^2/(4g_1^2) < k^2 < \infty$. This simple
example shows that although inclusion of the $a_1$-meson reduces the chance of
the kinetic energy of the theory becoming negative, it is not able to cure the
pathologies of the theory at the three-point level.

In general then the Hamiltonian associated with the simplest three-point
$\pi\rho a_1$ interactions {\it is not bounded from below} which is of course
unacceptable. This extends the results of reference [6] where the
$\pi\rho$ system was studied and where the energy of a topologically nontrivial
configuration was found to be unbounded from below when one includes only
three-meson couplings.

\bigskip

\bigskip

\centerline{\bfmagc{4. Four-point couplings.}}

\medskip

In order to cure the pathologies of the Lagrangian (9) we need to consider
higher-order terms. As we have seen, troubles emerge because the derivative
nature of vector-meson interactions produces singularities in the
kinetic part of the Hamiltonian. This suggests that one should analyse the
role of four-point couplings that are quadratic in time derivatives of the
pion field. The most general chiral Lagrangian at quartic order satisfying
$C$ and $P$ invariance, and leading to a Hamiltonian
that is at most quadratic in the momenta is:
$$\eqalign{
{\cal{L}}_{\pi\rho a_1}^{(4)}=&{1\over 8}\bigg\{g_4<[u_\mu,u_\nu]^2>+2g_5
<[u_\mu,u_\nu][A^\mu,u^\nu]>+2g_6\big(<[V_\mu,u_\nu]^2>\cr
&-<[V_\mu,u_\nu][V^\nu,u^\mu]>\big)+2g_7\big(<[A_\mu,u_\nu]^2>-<[A_\mu,u_\nu]
[A^\nu,u^\mu]>\big)\bigg\},\cr}\eqno(20)
$$
where we have introduced four new coupling constants $g_4,\ g_5,\ g_6,\ g_7$.
Amongst these terms one can recognise a local four-point pion vertex, the
so-called \lq\lq Skyrme term", as well as $\rho\rho\pi\pi$ and $a_1a_1\pi\pi$
vertices and a term contributing to the decay $a_1\to \pi\pi\pi$.

The potential energy ${\tilde H}_V$ of the theory reads now:
$$\eqalign{
\tilde{H}_V=H_V-{1\over 4}&\int d^3x\bigg\{g_4 [u_i,u_j]_k^2+2g_5[u_i,u_j]_k
[A_i,u_j]_k+2g_6([V_i,u_j]_k^2\cr
&-[V_i,u_j]_k[V_j,u_i]_k)+2g_7([A_i,u_j]_k^2-[A_i,u_j]_k[A_j,u_i]_k)\bigg\},
\cr}\eqno(21)
$$
where $H_V$ is the functional given by eq.~(10), and we use the tilde to denote
the corresponding quantity with four-point couplings included. The functional
form of the kinetic term, eq.~(11), is not modified by the inclusion of
four-point couplings: these are entirely contained in the corresponding
$\tilde{\vec\Gamma}$, $\tilde{\cal{P}}$ and $\tilde{\cal{A}}$. The expressions
for these quantities can be obtained from those at the three-point level by the
replacements $g_1^2 \to g_1^2-g_4, \ g_2^2 \to g_2^2-g_6, \ g_3^2 \to
g_3^2-g_7$ and $g_1g_3 \to g_1g_3-\displaystyle{{{g_5}\over 2}}$.

Turning to the energy of the charge-zero meson configuration defined in the
previous section, we note first that its potential energy is unaffected by the
new couplings and is still given by eq.~(14). The kinetic piece has the same
form as in eq.~(16) but with a new inertial function, $\tilde{\cal{I}}$. After
a tedious but straightforward calculation one finds:
$$\eqalignno{
\tilde{\cal{I}}={1\over{f^4\tilde{M}_1\tilde{M}_2}}
\bigg\{&f^4\tilde{M}_1\tilde{M}_2-8(g_2^2-g_6)V_i^2M_\rho^2\tilde{M}_2\cr
&+16\bigg[(g_2^2-g_6)^2\big(V_i^2(\partial_i F)^2
-(\partial_i F V_i)^2\big)\bigg]\tilde{M}_2\cr
&-8\bigg[(g_1^2-g_4)(\partial_i F)^2+(g_3^2-g_7)A_i^2-2\left(g_3g_1-{{g_5}
\over 2}\right)(\partial_i F A_i)\bigg]M_a^2 \tilde{M}_1\cr
&+16 \bigg[\bigg((g_1^2-g_4)(g_3^2-g_7)-\bigg(g_3g_1-{{g_5}\over 2}\bigg)^2
\bigg)(\partial_i F)^4&(22)\cr
&\qquad+(g_3^2-g_7)^2\big(A_i^2(\partial_i F)^2
-(A_i\partial_i F)^2\big) \bigg]\tilde{M}_1\bigg\},\cr}
$$
with $\tilde{M}_1=(1/f^2)\displaystyle{\big[2M_\rho^2-4(g_2^2-g_6)(\partial_i
F)^2\big]}$ and
$\tilde{M}_2=(1/f^2)\displaystyle{\big[2M_a^2-4(g_3^2-g_7)(\partial_i
F)^2\big]}$.

We are now in a position to find constraints on the couplings by requiring that
for any value of the classical profiles $\partial_i F,\ V_i,\ A_i$ the function
$\tilde{\cal{I}}$ is non-negative. To do this we consider three simplifying
cases where some of the fields vanish. These and their corresponding forms for
$\tilde{\cal{I}}$ are as follows:

\medskip

\noindent {\hskip 0.5cm} $\displaystyle{ a) \ \ \ \partial_i F=A_i=0 \ \ \ \ \
\Rightarrow \ \ \ \ \ \tilde{\cal{I}}_a=1-{4\over{f^2}}(g_2^2-g_6)V_i^2}$

\bigskip

\noindent {\hskip 0.5cm} $\displaystyle{ b) \ \ \ \partial_i F=V_i=0 \ \ \ \ \
\Rightarrow \ \ \ \ \ \tilde{\cal{I}}_b=1-{4\over{f^2}}(g_3^2-g_7)A_i^2}$
{\hfill} (23)
$$\eqalign{
c) \ \ \ V_i=A_i=0 \ \ \ \ \ \ \ \Rightarrow \ \ \ \ \ \tilde{\cal{I}}_c
=&\displaystyle{1\over{f^2\tilde{M}_2}}\bigg[2M_a^2-\big[4(g_3^2-g_7)
+8(g_1^2-g_4)\displaystyle{{M_a^2}\over{f^2}}\big](\partial_i F)^2\cr
&+\displaystyle{{16}\over{f^2}}\big[(g_1^2-g_4)(g_3^2-g_7)-(g_3g_1-{{g_5}
\over 2})^2\big](\partial_i F)^4\bigg].\cr}
$$
Requiring positiveness of $\tilde{\cal{I}}_{a,b,c}$ for all possible values
of the fields leads to the following constraints on the couplings constants:
$$\eqalign{
g_4 \  \ge& \ g_1^2\cr
g_6 \ \ge& \ g^2_2\cr
g_7 \  \ge& \ g_3^2\cr
(g_1^2-g_4)(g_3^2-g_7) \ \ge& \ (g_3g_1-{{g_5}\over 2})^2.\cr
}\eqno(24)
$$
The second and third inequalities follow immediately from requiring ${\cal
I}_{a,b}$ to be positive definite. They imply that both $\tilde{\cal{M}}_1$ and
$\tilde{\cal{M}}_2$ are also positive definite. For large amplitude fields the
quartic power $(\partial_i F)^4$ dominates over the quadratic one $(\partial_i
F)^2$ in the expression of ${\cal{I}}_c$. By demanding ${\cal{I}}_c$ to be
non-negative for these configurations one arrives at the fourth inequality. The
first inequality, which places a lower bound on the coefficient of the Skyrme
term was previously obtained in [6] from a Lagrangian with $\pi$- and
$\rho$-mesons. In the present case it results from combining the third and
fourth inequalities.

These conditions (24) show that the Skyrme term and other four-point
interactions are essential if the Hamiltonian is to be bounded from below. In
an effective theory where the spin-1 fields transform homogeneously they arise
as counterterms for the bad behaviour of the vector-meson
contributions. This is in sharp contrast to the approach of [3,4], where the
same Skyrme term emerges from the exchange of a very heavy $\rho$-meson.

The constraints we obtain here ensure that the kinetic energies of the specific
charge-zero meson configurations considered are bounded from below. Other
configurations, including ones with non-zero winding numbers, can also be
investigated but we have not found any which lead to more stringent
constraints on the couplings. We believe that our results are general for
any theory defined by a Lagrangian of the form (8, 9, 20).

\bigskip

\bigskip

\centerline{\bfmagc{5. Discussion}}

\medskip

To summarise our results so far: our investigation of classical nonperturbative
effects in low-energy chiral theories shows that the constraints (24), relating
three- and four-point couplings, must be satisfied for a consistent description
of the interactions between pions and spin-1 isovector mesons. We stress that
chiral symmetry is implemented nonlinearly in this approach and the vector
mesons are naturally assumed to transform homogeneously under chiral rotations.
The constraints arise from demanding that the Hamiltonian be bounded from
below. They do not depend on phenomenological ideas such as vector dominance.

Before trying to determine phenomenologically the various coupling constants in
this effective Lagrangian of pions, $\rho$'s and $a_1$'s, one might ask whether
there are any other constraints on them from first principles. For instance
another nonperturbative notion that one could invoke in this context is the
unitarity of the scattering matrix. This was studied in ref.~[10] for the
special case of the Lagrangian (8, 9) without the $a_1$ ($g_2=g_3=0$). Working
at tree-level, or leading order in a $1/N_c$ expansion, the authors of
ref.~[10] found  that further local interactions between the pions must be
added by hand  if the forward elastic $\pi\pi$ scattering amplitude is to obey
the Froissart bound [11]. These local interactions compensate for the most
divergent contribution  produced by $\rho$-exchange. In the three-flavor case
they have the form
$$\eqalign{
{\cal{L}}^{SU(3)}_{local}={{g_1^2}\over 8}\bigg\{&<\partial_\mu U
\partial^\mu U^\dagger>^2
+2 <\partial_\mu U^\dagger\partial_\nu U> <\partial^\mu U^\dagger
\partial^\nu U>\cr
&-6 <\partial_\mu U^\dagger\partial^\mu U \partial_\nu U^\dagger
\partial^\nu U > \bigg\},\cr}\eqno(25)
$$
where $U$ is an $SU(3)$ mapping. Converting this to our notation via the
relation  $\partial_\mu U U^\dagger=(1/i) u u_\mu u^\dagger$ and using
$\tau_a\tau_b=\delta_{ab}+i\epsilon_{abc}\tau_c$ to reduce it to the SU(2)
sector, we find that it is
$$\eqalign{
{\cal{L}}^{SU(2)}_{local}={{g_1^2}\over 8}  < [u_\mu,u_\nu]^2 >.\cr}\eqno(26)
$$
This is just the Skyrme term, but with a coefficient that is fixed by the
three-point coupling $g_1$. If one works at tree level, as in ref.~[10], the
$a_1$ does not contribute to $\pi\pi$ scattering, and so this value for the
four-point coupling is also appropriate to our more general $\pi\rho a_1$
theory. Hence imposing unitarity as in ref.~[10] leads to saturation of the
lower bound on $g_4$ in (24):
$$\eqalign{
g_4-g_1^2=0.\cr}\eqno(27)
$$
Combining this with the final constraint in (24), we obtain a similar relation
expressing the implications  of unitarity for the couplings of the {\it axial}
meson:
$$\eqalign{
g_5=2g_1g_3.\cr}\eqno(28)
$$
This relates the strength of the $a_1\to\pi\pi\pi$ decay to those of the
processes $\rho\to\pi\pi$ and $a_1\to\rho\pi$.

This saturation of two of our constraints in (24) follows from the assumption
that a single vector meson state contributes in the forward $\pi\pi$ scattering
amplitude -- an extreme version of vector dominance for strong interactions.
Realistically one also expects higher-mass vector mesons to contribute;
including them in the unitarity argument would require additional terms of the
form (6). The coefficient of the Skyrme term would not then be given in terms
of the $\rho\pi\pi$ coupling $g_1$ alone. It would continue to satisfy the
first of our inequalities (24), but not the equality (27). As shown in
Ref.~[10] the value for this coefficient determined assuming
$\rho$-meson dominance agrees well with that from chiral perturbation theory
[12]. This suggests that the vector dominance assumption holds to a reasonable
accuracy. We speculate that a similar assumption of dominance of a single
resonance may also hold in the axial-vector channel, leading to saturation of
the remaining constraints in (24). In this case our lagrangian would simplify
to:
$$\eqalign{
{\cal{L}}_{\pi\rho a_1}=&{{f^2}\over 4}<u_\mu u^\mu>+{{M_{\rho}^2}\over 2}
<V_\mu V^\mu>+{{M_a^2}\over 2}<A_\mu A^\mu>\cr
-&{1\over 4}<\bigg(V_{\mu\nu}+{i\over{\sqrt{2}}}\big(g_1[u_\mu,u_\nu]
+g_3([A_\mu,u_\nu]-[A_\nu,u_\mu])\big)\bigg)^2>\cr
-&{1\over 4}<\bigg(A_{\mu\nu}
+{i\over{\sqrt{2}}}g_2([V_\mu,u_\nu]-[V_\nu,u_\mu])\bigg)^2>.\cr}\eqno(28)
$$
This constitutes an effective lagrangian describing  the strong interactions of
$\pi\rho a_1$ mesons with a minimal number of free coupling constants. It is
amusing to  note that in this case the transformation matrix relating the pion
time derivative and its momentum still has the form of the original  nonlinear
sigma model. This new lagrangian is the simplest one compatible with chiral
symmetry and leading to a hamiltonian which is free of pathologies. We believe
that it should be regarded as the starting point for any extension of chiral
perturbation theory [12] to the resonance region.

As a future prospect, and in connection with baryon physics, let us mention
that all past attempts to build topological solitons of the $\pi\rho a_1$
system have failed: the solitons of previously proposed lagrangians have
been shown to be generically unstable [13,14]. We would like to stress here
that these attempts were only based on massive Yang-Mills or hidden gauge
symmetry assumptions. It would now be very interesting to investigate the issue
of soliton stability in the alternative framework described here for $\pi\rho
a_1$  physics.

Finally it will be important to compare the predictions of the various
treatments of $\rho$ and $a_1$ mesons for processes like
$\rho\to\pi\pi\pi\pi$. Accurate measurements of these at DA$\Phi$NE [15] could
provide a stringent test of effective theories including vector mesons [16].

\bigskip

\centerline{\bfmagc Acknowledgements}

\bigskip

We are grateful to Bachir Moussallam for communication of his results
on the asymptotic behaviour of three-point functions in low energy effective
lagrangians. DK benefitted from the remarks and encouragements of Robert Vinh
Mau during the early stages of this work. MCB is grateful to the SERC for an
Advanced Fellowship and DK acknowledges support from the SERC.

\eject

\bigskip

\centerline{\bfmagc Appendix: Construction of $H_T$}

\bigskip

In this note we present details of the derivation of the kinetic part of the
secondary Hamiltonian density which appears in eq.~(11). We display here the
results for the three-point theory defined in section 3 by the Lagrangian
${\cal{L}}_{\pi\rho a_1}^{(2)}+{\cal{L}}_{\pi\rho a_1}^{(3)}$. The structures
which appear in the Hamiltonian for the theory with four-point interactions
are exactly the same and so the corresponding expression can be obtained by
appropriate substitutions of combinations of the couplings, as described in
section 4.

We first build the primary Hamiltonian. The conjugate momenta ${\vec \phi,\
\vec \pi_i,\ \vec \chi_i}$ for the fields $\vec F,\ \vec V_i,\ \vec A_i$ can be
found in the usual way by differentiating the Lagrangian with respect to the
time derivatives of the fields. Inverting this relation yields:
$$\eqalign{
\dot{\vec F}=&{\cal{A}}^{-1}(\vec\phi- {\cal{B}}_i \vec\pi_i - {\cal{C}}_i
\vec\chi_i -\vec \theta)\cr
\dot{\vec V_i}=&{-{\vec\pi_i}\over 2}+{\cal{B}}_i^T{\cal{A}}^{-1}
(\vec\phi- {\cal{B}}_j \vec\pi_j - {\cal{C}}_j
\vec\chi_j -\vec \theta)+\vec \zeta^V_i\cr
\dot{\vec A_i}=&{-{\vec\chi_i}\over 2}+{\cal{C}}_i^T{\cal{A}}^{-1}(\vec\phi-
{\cal{B}}_j \vec\pi_j - {\cal{C}}_j
\vec\chi_j -\vec \theta)+\vec \zeta^A_i,\cr}\eqno(A.1)
$$
where the script capital letters denote $3 \times 3$ matrices acting on
isospin vectors and the superscript $T$ denotes transposition.
The matrices in these equations are:
$$\eqalign{
{\cal{A}}={\cal{A}}^T=&f^2{\cal{G}}-4(g_1{\cal{N}}_i-g_3{\cal{Q}}_i^{A})^T
(g_1{\cal{N}}_i-g_3{\cal{Q}}_i^{A})-4g^2_2({\cal{Q}}_i^{V})^T{\cal{Q}}_i^{V}\cr
{\cal{B}}^T_i=&-{1\over 2}\bigg\{2{\cal{M}}_i^{V}-{4\over{\sqrt{2}}}
(g_1{\cal{N}}_i-g_3{\cal{Q}}_i^{A})\bigg\}\cr
{\cal{C}}^T_i=&-{1\over 2}\bigg\{2{\cal{M}}_i^{A}+{4\over{\sqrt{2}}}
g_2{\cal{Q}}_i^{V}\bigg\},\cr}\eqno(A.2)
$$
with the definitions (all indices label isospin, except $i$ and $j$
which we use for space components):
$$\eqalign{
({\cal{G}})_{ab}=&\hat F_a\hat F_b+{{\sin^2 F}\over {F^2}}(\delta_{ab}-
\hat F_a\hat F_b)\cr
({\cal{M}}_i^V)_{ab}=&{{\partial (V_{0i})_a}\over{\partial(\partial_0 F_b)}}=
-{{2\sin^2(F/2)}\over {F^2}}\big[\delta_{ab}
(\vec F.\vec V_{i})- F_a (V_{i})_b\big]\cr
({\cal{M}}_i^A)_{ab}=&{{\partial (A_{0i})_a}\over{\partial(\partial_0 F_b)}}=
-{{2\sin^2(F/2)}\over {F^2}}\big[\delta_{ab}
(\vec F.\vec A_{i})- F_a (A_{i})_b\big]\cr
({\cal{N}}_i)_{ab}=&\epsilon_{pqa}\partial_i F_r \displaystyle{(\sqrt
{{\cal{G}}})}_{pb}\displaystyle{(\sqrt{{\cal{G}}})}_{qr}\cr
({\cal{Q}}^V_i)_{ab}=&-\epsilon_{pqa}(V_i)_p\displaystyle{(\sqrt
{{\cal{G}}})}_{qb}\cr
({\cal{Q}}^A_i)_{ab}=&-\epsilon_{pqa}(A_i)_p\displaystyle{(\sqrt
{{\cal{G}}})}_{qb}.\cr}\eqno(A.3)
$$
The isospin vectors in (A.1) are linear functions of the dependent variables
$\vec V_0$ and $\vec A_0$ and are given by:
$$\eqalign{
\theta_k=&2i g_3 (g_1{\cal{N}}_i-g_3{\cal{Q}}_i^{A})_{mk}
[A_0,u_i]_m-2i g_2^2 ({\cal{Q}}_i^{V})_{mk} [V_0,u_i]_m\cr
(\zeta^V_i)_k=&(\nabla_i V_0)_k-{{ig_3}\over{\sqrt{2}}}[A_0,u_i]_k\cr
(\zeta^A_i)_k=&(\nabla_i A_0)_k-{{ig_2}\over{\sqrt{2}}}[V_0,u_i]_k.\cr}
\eqno(A.4)$$

The Hamiltonian is given by the following Legendre transformation:
$$\eqalign{
H=\int d^3 x \bigg\{\vec\phi\dot{\vec F}-\vec\pi_i \dot{\vec V_i}-\vec\chi_i
\dot{\vec A_i}-
{\cal{L}}_{\pi\rho a_1}^{(2)}-{\cal{L}}_{\pi\rho a_1}^{(3)}\bigg\}.\cr}
\eqno(A.5)$$
It can be split into two pieces $H=H_T '+H_V$, where the primary \lq\lq
kinetic" energy is $H_T '$ and the \lq\lq potential" energy is $H_V$. The form
of the latter is given in eq.~(10). The primary \lq\lq kinetic" energy, which
is our concern here, contains all the pieces that depend on the conjugate
momenta or involve $\vec V_0,\ \vec A_0$:
$$\eqalign{
H_T '=\int d^3 x \bigg\{&
{1\over 2} ( \vec\phi- \vec\pi_i {\cal{B}}^T_i - \vec\chi_i
{\cal{C}}^T_i-\vec\theta)
{\cal{A}}^{-1} (\vec\phi - {\cal{B}}_i \vec\pi_i - {\cal{C}}_i
\vec\chi_i-\vec\theta)
+{{\vec\pi_i^2}\over 4} + {{\vec\chi_i^2}\over 4}\cr
&-\vec\pi_i\vec\zeta_i^V-\vec\chi_i\vec\zeta_i^A-
M_\rho^2 \vec V_0^2 - M_a^2 \vec A_0^2- {{g^2_3}\over{2}}[A_0,u_i]_k^2-
{{g^2_2}\over{2}}[V_0,u_i]_k^2\bigg\}.\cr}\eqno(A.6)
$$
As can be seen this involves a nontrivial mixing between the pion conjugate
momentum and the time components of both vector fields and axial fields. This
is in contrast to the minimal $\pi\rho$ case studied in [6] where such a mixing
does not occur. This feature complicates the determination of the kinetic part
of the Hamiltonian as a function of the independent dynamical variables only,
as is needed for proper quantisation.

We wish to eliminate the dependent variables $\vec V_0,\ \vec A_0$ from the
expression for the energy. For this purpose it proves convenient to rewrite the
primary $H_T '$ functional in the following suggestive form:
$$\eqalign{
H_T '=\int d^3 x \bigg\{
{1\over 2} \vec\Phi   {\cal{A}}^{-1} \vec\Phi  +{{\vec\pi_i^2}\over 4}
+ {{\vec\chi_i^2}\over 4}
-\bigg[\vec\Gamma\vec\Delta+{1\over 2}\vec\Delta \ {\cal{P}} \
\vec\Delta\bigg]  \bigg\},\cr}\eqno(A.7)
$$
where we have introduced the notation $\vec\Phi= \vec\phi- {\cal{B}}_i
\vec\pi_i - {\cal{C}}_i \vec\chi_i$ for simplicity and done some integration by
parts so that the gradients of $\vec V_0,\ \vec A_0$ do not appear in $H_T'$.
The symbols $\vec\Delta$, $\vec\Gamma$ denote two-component vectors in the
space spanned by $\vec V_0$ and $\vec A_0$. Their expressions read:
$$\eqalign{
\vec\Gamma=&\pmatrix{\vec\alpha\cr\vec\beta\cr},  \ \ \vec\Delta=
\pmatrix{\vec V_0\cr\vec A_0\cr}
\cr}\eqno(A.8)
$$
with
$$\eqalign{
\alpha_k=&-(\nabla_i\pi_i)_k+{{ig_2}\over\sqrt{2}}[\chi_i,u_i]_k+4g_2^2
({\cal{Z}}^V{\cal{A}}^{-1})_{kp}\Phi_p\cr
\beta_k=&-(\nabla_i\chi_i)_k+{{ig_3}\over\sqrt{2}}[\pi_i,u_i]_k+4
\bigg[(g_3^2{\cal{Z}}^A-g_3g_1{\cal{Z}}^\pi){\cal{A}}^{-1}\bigg]_{kp}\Phi_p\cr
{\cal{Z}}^V_{km}=& ({\cal{Q}}^V_j)_{am} \epsilon_{kra} (u_j)_r,
\ {\cal{Z}}^A_{km}=({\cal{Q}}^A_j)_{am}
\epsilon_{kra} (u_j)_r, \ {\cal{Z}}^\pi_{km}=({\cal{N}}_j)_{am}
\epsilon_{kra} (u_j)_r.\cr}
\eqno(A.9)
$$
In eq.~(A.7), ${\cal{P}}$ is a $2\times 2$ matrix acting on these vectors:
$$\eqalign{
{\cal{P}}=&\pmatrix{D&E\cr E^T&H\cr}.}\eqno(A.10)
$$
The isospin content of the matrix elements $D, E, H$ is:
$$\eqalign{
D_{km}=&2M_\rho^2\delta_{km}-4g_2^2(\vec
u_i^2\delta_{km}-u_i^k u_i^m)-16\bigg[g_2^2{\cal{Z}}^V{\cal{A}}^{-1}g_2^2
({\cal{Z}}^V)^T\bigg]_{km}\cr
H_{km}=&2M_a^2\delta_{km}-4g_3^2(\vec
u_i^2\delta_{km}-u_i^k u_i^m)-16\bigg[(g_3^2{\cal{Z}}^A-g_3g_1{\cal{Z}}^\pi)
{\cal{A}}^{-1}(g_3^2{\cal{Z}}^A-g_3g_1{\cal{Z}}^\pi)^T\bigg]_{km}\cr
E_{km}=&-16\bigg[g_2^2{\cal{Z}}^V{\cal{A}}^{-1}
(g_3^2{\cal{Z}}^A-g_3g_1{\cal{Z}}^\pi)^T\bigg]_{km}.\cr}\eqno(A.11)
$$

In the canonical formalism the conservation in time of the primary constraints
of the theory $\vec\pi_0=\vec\chi_0=0$ is used in order to eliminate the
$\vec\Delta$-dependence of the hamiltonian. This conservation law amounts to
the vanishing of the Poisson brackets of these momenta with $H_T'$. This leads
to six equations relating the constrained time components of the $\rho$ and the
$a_1$ fields to other fields and their conjugate momenta. In our compact matrix
notation these take the following simple form:

\medskip

\noindent {\hskip 3cm} $\bigg
\{\displaystyle{\pmatrix{\vec\pi_0\cr\vec\chi_0\cr}}, H_T '\bigg\}=
\vec\Gamma+{\cal{P}}\vec\Delta=0$, \hfill (A.12)

\medskip

\noindent where $\{ \ , \ \}$ denotes the Poisson bracket. One can check that
for $g_2=g_3=0$ this reduces to the covariant form of Gauss's law for
massive vector fields [6]: $V_0=(1/2M_\rho^2)\nabla_i\pi_i$.

If we suppose the matrix ${\cal{P}}$ to be invertible, $\vec\Delta$ can
finally be removed from the Hamiltonian using eq.~(A.12) to leave us with a
functional of the independent dynamical variables $\vec F,\ \vec V_i,\ \vec
A_i,$ $\vec\phi,\ \vec\pi_i,\ \vec\chi_i$ only. This is the secondary kinetic
energy $H_T$ that we study in the text:
$$\eqalign{
H_T=\int d^3 x \bigg\{
{1\over 2} \vec\Phi   {\cal{A}}^{-1} \vec\Phi  +{{\vec\pi_i^2}\over 4} +
{{\vec\chi_i^2}\over 4}
+{1\over 2} \vec\Gamma {\cal{P}}^{-1} \vec\Gamma \bigg\},\cr}\eqno(A.13)
$$
where the inverse of ${\cal{P}}$ can be written
$$\eqalign{
{\cal{P}}^{-1}=\pmatrix{(D-EH^{-1}E^T)^{-1}& \ \ -(D-EH^{-1}E^T)^{-1}EH^{-1}\cr
\ \ \cr
-(H-E^TD^{-1}E)^{-1}E^TD^{-1}&(H-E^TD^{-1}E)^{-1}\cr}.}\eqno(A.14)
$$

\eject

\centerline{\bfmagd References}

\bigskip

\item{[1]}S. Weinberg, Phys. Rev. 166 (1968) 1568;
I. Gerstein, R. Jackiw, B. Lee and S. Weinberg, Phys. Rev. D3 (1971) 2486;
A. Salam and J. Strathdee, Phys. Rev. D2 (1970) 2869.
\medskip
\item{[2]}G. 't Hooft, Nucl. Phys. B79 (1974) 276.
\medskip
\item{[3]}J. J. Sakurai, {\it Currents and Mesons}, University of Chicago
Press, Chicago 1969;
J. Schwinger, {\it Particles and Sources}, Clarendon Press, Oxford, UK, 1969.
\medskip
\item{[4]}M. Bando, T. Kugo and K. Yamawaki, Phys. Rep. 164 (1988) 217.
\medskip
\item{[5]}G. Ecker, J. Gasser, A. Pich and E. de Rafael, Nucl. Phys. B321
(1989) 311.
\medskip
\item{[6]}D. Kalafatis, Phys. Lett. B313 (1993) 115.
\medskip
\item{[7]}S. Coleman, J. Wess and B. Zumino, Phys. Rev. 177 (1969)
2239; C. G. Callan, S. Coleman, J. Wess and B. Zumino, {\it ibid} 2247.
\medskip
\item{[8]}J. Bijnens, C. Bruno and E. de Rafael, Nucl. Phys. B390 (1993) 501;
E. Pallante and R. Petronzio, Nucl. Phys. B396 (1993) 205;
E. Pallante and R. Petronzio, Roma preprint ROM2F 93/37.
\medskip
\item{[9]}B. Moussallam, private communication.
\medskip
\item{[10]}G. Ecker, J. Gasser, H. Leutwyler, A. Pich and E. de Rafael, Phys.
Lett.  B223 (1989) 425.
\medskip
\item{[11]}M. Froissart, Phys. Rev. 123 (1961) 1053;
A. Martin, Phys. Rev. 129 (1963) 1432.
\medskip
\item{[12]}J. Gasser and H. Leutwyler, Ann. Phys. (NY) 158 (1984) 142;
Nucl. Phys. B250 (1985) 465, 517, 538.
\medskip
\item{[13]}H. Forkel, A. D. Jackson and C. Weiss, Nucl. Phys. A526 (1991) 453.
\medskip
\item{[14]}Z. F. Ezawa and T. Yanagida, Phys. Rev. D33 (1986) 247 ;
J. Kunz and D. Masak, Phys. Lett. B179 (1986) 146.
\medskip
\item{[15]}{\it Proposal for a Phi factory}, Laboratori Nazionali di
Frascati-INFN report LNF-90/031 (R) (april 1990).
\medskip
\item{[16]}A. Bramon, A. Grau and G. Pancheri, Phys. Lett. B317 (1993) 190.
\bye